# Experimental Observation of Berry Phases in Optical Möbius-strip Microcavities

Jiawei Wang[1,2,3,11], Sreeramulu Valligatla[1,11], Yin Yin[1,4], Lukas Schwarz[1], Mariana Medina-Sánchez[1], Stefan Baunack[1], Ching Hua Lee[5], Ronny Thomale[6], Shilong Li[7], Vladimir M. Fomin[1,8], Libo Ma[1] and Oliver G. Schmidt[2,9,10]

[1] Institute for Integrative Nanosciences, Leibniz IFW Dresden, Helmholtzstr. 20, D-01069 Dresden, Germany

[2] Research Center for Materials, Architectures and Integration of Nanomembranes (MAIN), Technische Universität Chemnitz, Rosenbergstr. 6, Chemnitz, D-09126 Germany

[3] School of Electronic and Information Engineering, Harbin Institute of Technology (Shenzhen), Shenzhen, 518055 China

[4] School of Materials Science and Engineering, Jiangsu University, 212013 Zhenjiang, China

[5] Department of Physics, National University of Singapore, 117551 Singapore

[6] Institute for Theoretical Physics and Astrophysics, University of Würzburg, Am Hubland, D-97074 Würzburg, Germany

[7] Light-Matter Interactions for Quantum Technologies Unit, Okinawa Institute of Science and Technology Graduate University, Onna, Okinawa 904-0495, Japan

[8] Laboratory of Physics and Engineering of Nanomaterials, Department of Theoretical Physics, Moldova State University, str. A. Meteevici 60, MD-2009, Chişinău, Republic of Moldova

[9] Material Systems for Nanoelectronics, Chemnitz University of Technology, Str. der Nationen 62, Chemnitz, D-09111 Germany

[10] Nanophysics, Dresden University of Technology, Haeckelstr. 3, Dresden, D-01069, Germany

[11] These authors contributed equally: Jiawei Wang, Sreeramulu Valligatla

Corresponding author: S. L. (shilong.li@oist.jp); L. B. M. (l.ma@ifw-dresden.de)

**The Möbius strip, as a fascinating loop structure with one-sided topology, provides a rich playground for manipulating the non-trivial topological behavior of spinning particles, such as electrons, polaritons, and photons in both real and parameter spaces. For photons resonating in a Möbius-strip cavity, the occurrence of an extra phase, known as Berry phase, with purely topological origin is expected due to its non-trivial evolution in the parameter space. However, despite numerous theoretical investigations, characterizing optical Berry phase in a Möbius-strip cavity has remained elusive. Here we report the experimental observation of Berry phase generated in optical Möbius-strip microcavities. In contrast to theoretical predictions in optical, electronic, and magnetic Möbius-topology systems where only Berry phase $\pi$ occurs, we demonstrate that variable Berry phase smaller than $\pi$ can be acquired by generating elliptical polarization of resonating light. Möbius-strip microcavities as integrable and Berry−phase−programmable optical systems are of great interest in topological physics and emerging classical or quantum photonic applications.**

Berry phase (also called "geometric phase"), as a non-integrable phase factor, originates from a non-trivial evolution of a physical system in a parameter space[1], playing a fundamental role in various fields ranging from condensed matter physics[2,3], acoustics[4], high-energy physics[5], cosmology[6], quantum information[7] to optics[8]. In optics, Berry phase can be acquired by the non-trivial evolution of either the polarization state or the wave vector in its corresponding parameter space with purely topological origin[8-13]. The investigation of manipulating polarization states has been pioneered by S. Pancharatnam, by which the generated geometric phase is now also called Pancharatnam-Berry phase[1,10]. The generation of Berry phase based on wave vector evolution has been explored by recording polarization rotations in open light paths of helical[14-17] or out-of-plane curvilinear waveguides[18]. Notably, in recent years, the generation of topologically protected Berry phase has also been studied in specially designed 3D optical microcavities with light circulating along a closed path, such as asymmetric whispering gallery mode microcavities[19], out-of-plane microrings[20], and microrings with embedded angular scatterers[21], showing great potential for on-chip integrated topological photonic devices.

The Möbius strip[22] as a fascinating loop structure is well-known for its one-sided topology, which also symbolizes the topological twist of the band structure in topological insulators[23]. The Möbius topology plays important roles in multiple disciplines ranging from the generation of symmetry-breaking Möbius soliton modes in magnetic medium[24] to the extraordinary behavior of electronic waves in Möbius aromaticity[25] and twisted semiconductor strips[26], and anomalous plasmon modes formed in metallic Möbius nanostructures[27]. To impose the Möbius topology on photons, the optical field was twisted by liquid crystal q-plates as cavity-free systems[28]. Nevertheless, investigating topological phenomena of light resonating in a real Möbius-structured cavity is highly sought-after. Very recently, a Möbius-strip cavity made of a twisted dielectric strip has been explored as a platform for the investigation of non-Euclidean optics[29]. However, optical Berry phase, as the key topological phenomenon, has not been experimentally revealed, though the existence of Berry phase in a Möbius cavity has been theoretically predicted a long time ago[30,31]. Theoretical studies have shown the occurrence of Berry phase $\pi$ in an ideal Möbius-strip cavity, resulting in constructive self-interference of half-integer number modes, i.e. accommodating half-integer numbers of wavelengths, in a closed-path trajectory[30]. The half-integer number of wavelengths, which is due to a purely topological origin, contradicts the well-known resonant condition in conventional optical or plasmonic cavities.

The topological behavior of circulating light in Möbius-strip waveguiding systems has not yet been explored experimentally, which prevents gaining new insights into observable optical phenomena for topology-based signal processing and communications. In the present work, we report the experimental observation of optical Berry phases occurring in Möbius-strip microcavities with tailored cross-sectional geometry. In contrast to previous theoretical predictions, where only phase $\pi$ occurs, here we observe and reveal programmable Berry phases ranging from $\pi$ to 0 for the resonant light waves with linear to elliptical polarization in carefully designed Möbius-strip resonators. As the quantum holonomy, Berry phase generated in the compact optical Möbius system is particularly promising for geometric quantum mechanics and its applications, like simulation, metrology, sensing, and computation.

**Results**

**Principles.** The discussion starts with the optical polarization states in an ideal Möbius strip, in which the strip thickness $T$ is much smaller than the strip width $W$ (Fig. 1). Considering linearly polarized light resonating in the Möbius-strip waveguide, the optical electric field is guided and forced to remain in the plane of the twisted strip. Such a twisted strip waveguide functions similarly to some free-space optical components, such as a half-wave plate[1] or a Dove prism[32], which affects the orientation of the polarization states. As a result, the polarization continuously re-orients along the twisted strip during the propagation, which we term the "in-plane (IP)" mode. The IP mode represents an adiabatic cyclic evolution of linearly polarized light in a smoothly curved waveguide. The acquired phase factor of the light wave can be divided into two parts, the dynamic phase and Berry phase (alternatively called geometric phase). The dynamic phase reflects the system's evolution in time, which is determined by the optical path and the system's curvature. In contrast, Berry phase memorizes the evolution path in the parameter space, which is independent of the dynamic phase. Using a phenomenological model, the occurrence of Berry phase in a Möbius strip can be directly visualized by the parallel transport of a vector along the twisted strip. For reference, we investigate a comparable 3D ring cavity (termed "curved strip") with the same curvature as that of a Möbius strip but without the Möbius topology (see Supplementary Fig. S1 and Supplementary Note 1). In contrast to the one-sided Möbius strip with a single surface, the curved strip is topologically identical to conventional microring cavities. As shown in Fig. 1, although the vector is kept in parallel at each local position during the transport, the vector flips by an angle $\pi$ after a full trip around the Möbius strip, while there is no such a flip in the curved strip.

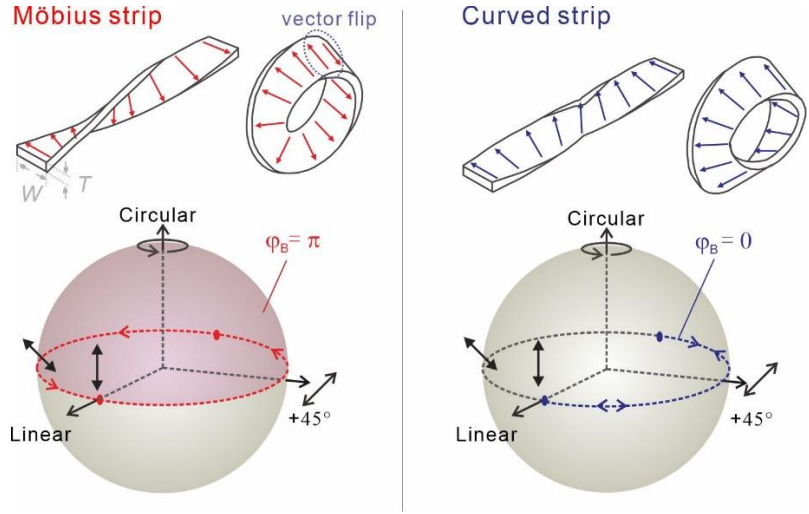

**Fig. 1 | Berry phase occurring in Möbius- and curved-strip microcavities.** Upper panels: parallel transport of a vector along Möbius- and curved-strip cavities leads to a vector flip (occurrence of Berry phase π, see dashed violet ellipse) and vector match (no Berry phase), respectively. Bottom panels: Corresponding vector transport evolution on the Poincaré sphere with/without solid angle for the Möbius/curved strips.

For the adiabatic cyclic evolution of a degenerate physical system, there are usually two analytical ways to quantify the occurring Berry phase via the corresponding solid angle in the parameter space. One way is determined by the wave vector $k$, which forms a sphere in the momentum space. Berry phase is equal to the solid angle enclosed by the trace of the wave vector at the origin of the momentum space[14-16]. The other way is related to the wave polarization state vector, which spans the Poincaré sphere. Berry phase is equal to half of the solid angle enclosed by the loop of the polarization vector at the origin of the Poincaré sphere[11,13].

For linear polarization, the polarization state can be described by $|s\rangle = \frac{1}{\sqrt{2}}(|+\rangle + |-\rangle)$, where $|+\rangle$ and $|-\rangle$ are right and left circular bases. The continuous variation of the

polarization orientation can be visualized by a closed loop along the equator of the Poincaré sphere (see Fig. 1, left panel), resulting in a solid angle $\Omega = 2\pi$. Hence, Berry phase as large as a half the solid angle $\Omega/2 = \pi$ is generated for the right and left circular polarization bases as $|s'\rangle = \frac{1}{\sqrt{2}}(e^{i\pi}|+\rangle + e^{-i\pi}|-\rangle)$ [11-13]. For optical resonances in the curved-strip cavities, the trajectory of the polarization evolution on the Poincaré sphere is topologically trivial and thus does not generate any Berry phase, as illustrated in Fig. 1. The similar photon propagation trajectory and propagation constant render the curved-strip cavity a perfect reference for our study of quantifying Berry phase.

**Light ellipticity in Möbius strips.** Two-photon polymerization-based direct laser writing was employed to fabricate dielectric Möbius- and curved-strip microrings using negative photoresist IP-Dip (see *Methods* and Supplementary Fig. S2). The polymerized IP-Dip is transparent from visible to near-infrared spectral range, and thus is suitable for supporting optical resonances in this spectral range. Figure 2a and b display scanning electron microscopy (SEM) images of Möbius- and curved-strip cavities with the identical design parameter $T/W \sim 0.33$ ($T = 0.9$ μm).

To understand the behavior of light propagation and resonances in all-dielectric Möbius microcavities, 3D numerical simulations based on the finite-element method were carried out (see *Methods*). Figure 2b and d present the calculated mode profiles in typical Möbius and curved strips with $T/W = 0.33$ ($T = 0.9$ μm), respectively. The electric field orientation of the resonant light rotates along the twisted strips, which can be seen from the magnified images at the horizontal and vertical sites of the Möbius and curved strips. In the Möbius-strip cavity, an odd number of antinodes ($N = 115$) is calculated implying the constructive interference with a half-integer mode number ($M = N/2 = 57.5$) of wavelengths.

The constructive interference is off-phase in the case of half-integer numbers of wavelengths, and the phase mismatch is precisely compensated by the presence of Berry phase π, which leads to a 180° wave-flip. In the curved-strip cavity, the even number ($N = 116$) of antinodes corresponds to the conventional constructive interference with an integer number ($M = 58$) of wavelengths.

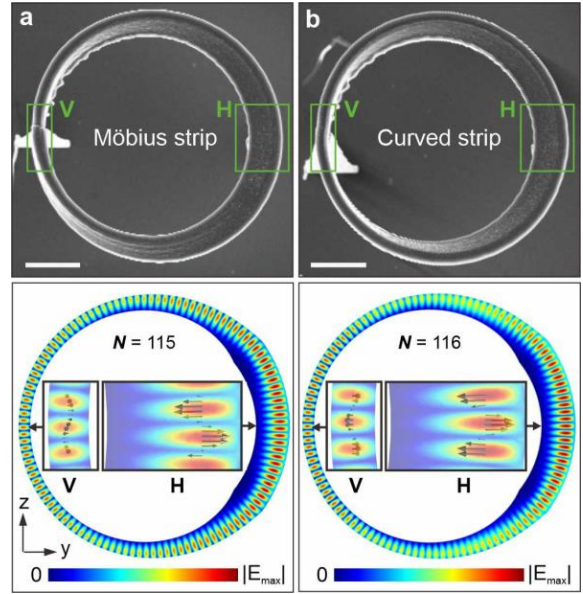

**Fig. 2 | Optical resonant modes with/without Berry phase in Möbius/curved-strip cavities.** SEM images of fabricated Möbius- (**a**) and curved-strip (**b**) cavities of the same size. Scale bar: 5 μm. Simulated optical mode profiles indicate that the polarization orientation of the resonant light rotates along the strip structure generating Berry phase. The presence/absence of Berry phase leads to electric field patterns with odd/even numbers of antinodes (cases of $N=115/116$ are shown as examples) in the Möbius-/curved-strip microcavities. Insets: Magnified images of the electric field at the vertical (V, where the strip is perpendicular to the substrate) and horizontal (H, where the strip is in parallel with the substrate) sites. Black arrows indicate the local polarization revealing the rotation of the electric field orientation along the cavities.

As a key approximation in an ideal Möbius strip, the strip thickness $T$ is assumed to be much smaller than the wavelength of the considered light in the waveguiding medium ($\lambda/n$), so that the optical field is strictly confined within the strip during propagation. In reality, the thickness of a Möbius-strip cavity can only be shrunk to a finite value due to the need for sufficient optical confinement and low-loss light propagation. When $T$ is comparable to $W$, the twisted cross-section does not provide rigorous in-plane confinement of the optical electric field. This leads to two competing effects. On the one hand, due to the electromagnetic inertia, the electric field tends to maintain its orientation as it propagates along the strip. On the other hand, the tilted cross-section forces the electric field to rotate along the twisted strip. As a result, the main electric field orientation tilts away from the strip plane, generating in-plane and out-of-plane components (see Fig. 3a) with a phase retardation due to the different effective refractive indices for the electric-field components in and perpendicular to the strip plane (see Supplementary Fig. S3 and Supplementary Note 1). As such, an elliptical polarization is formed in the cross-section of the Möbius-strip cavity, which is illustrated in Fig. 3a and b.

The resonant modes were characterized by measuring transmission spectra using an evanescently-coupled tapered nanofiber as a widely adopted approach for near-field delivery and probe of lightwaves (see *Methods*, Supplementary Fig. S4 and Supplementary Note 2)[33]. The polarization states of the resonant light in the Möbius- and curved-strip microcavities were examined by tuning the input optical polarization step by step. For the Möbius-strip cavity with T/W ~ 0.67, the measurement results in an elliptical polarization state as presented in Fig. 3c. This is in sharp contrast to the linear polarization state for the Möbius-strip cavity with T/W ~ 0.33. The ellipticity as a function of $T/W$ was further studied by numerical simulations (see Fig. 3d). The resonant light is almost linearly polarized when $T/W < 0.5$, and turns elliptically polarized when $T/W > 0.5$. A further increase in $T/W$ leads to a "torus-like"-ring structure (i.e., $T/W \sim 1$). In this case, the optical electric field is not forced anymore to rotate along the structure with a square-shaped core (see Supplementary Fig. S6 and Supplementary Note 4), and therefore the light does not undergo any non-trivial topological evolution.

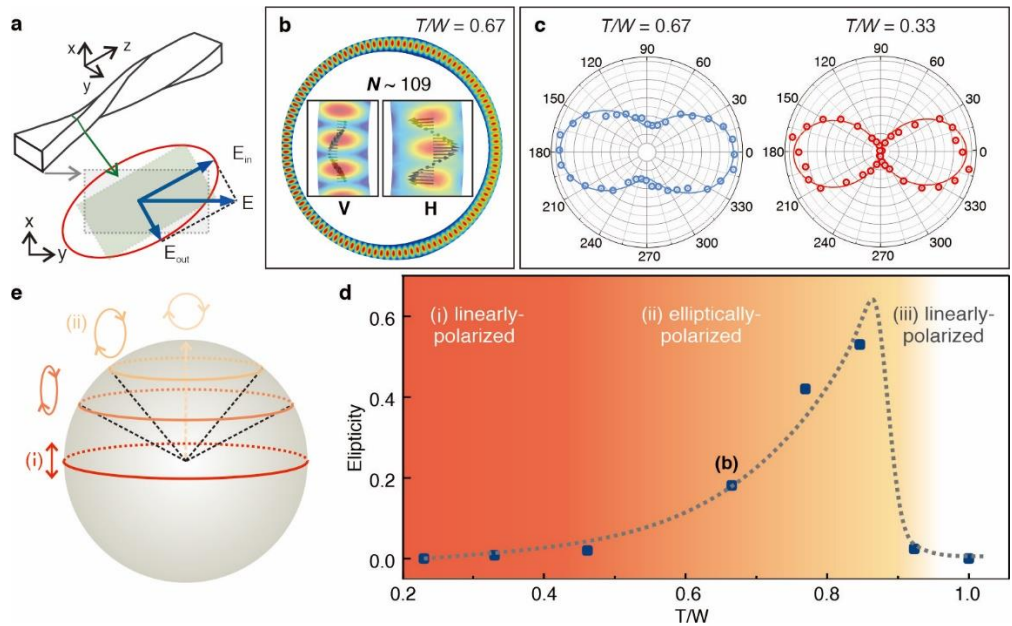

**Fig. 3 | Variable ellipticity of resonant modes generated in Möbius-strip cavities. a,** Linearly-polarized light propagating along a twisted strip leads to the existence of in-plane and out-of-plane components for the formation of an elliptical polarization state. **b,** Cross-sectional views of the electric field amplitude distributions for $T/W = 0.67$. Insets: magnified images of the electric field at the vertical and horizontal sites. Black arrows indicate the local polarization. **c,** Polarization-resolved measurements for Möbius strips with $T/W = 0.33$ and $0.67$. Dots: measured data. Lines: sinusoidal fits. **d,** Summarized light ellipticity extracted from the simulated modes at the horizontal site of Möbius strips as a

function of *T*/*W*. **e,** For simplicity, a cyclic evolution of the polarization state vector on a Poincaré sphere is used to describe the optical resonances in the Möbius strip, indicating that the occurring Berry phase directly depends on the ellipticity (represented by lines with arrows).

The evolution of elliptically polarized resonant light projects a loop on the Poincaré sphere away from the equator depending on its ellipticity (see Fig. 3e). Considering the solid angle $\Omega$ changing between $2\pi$ and 0, Berry phase $\varphi_B$ is generated for the right and left circular polarization bases $|s'\rangle = \frac{1}{\sqrt{2}}(e^{i\varphi_B}|+\rangle + e^{-i\varphi_B}|-\rangle)$. Therefore, by engineering the polarization ellipticity, one can develop a strategy to generate a variable Berry phase affecting both the eigenvalues and eigenstates of Möbius-strip cavities. For *T*/*W* ~1, the unmodified and non-rotating polarization state leads to the absence of Berry phase, indicating a negligible Möbius-topology effect on the resonant light. More details of Berry phase involved constructive interferences in a Möbius strip are further described in Supplementary Note 5.

**Manipulating Berry phases.** Controlling the cross-sectional dimension of a Möbius waveguiding system offers great flexibility in supporting optical resonances of elliptically-polarized light and hence manipulating the acquired Berry phases. The acquired Berry phases are summarized in Fig. 4a by comparing the resonant modes measured from pairs of Möbius- and curved-strip cavities with various *T*/*W* (see Methods). In the case of *T*/*W* = 1, no wavelength offsets are observed between the Möbius- and curved-strip cavities, indicating the absence of the extra phase, i.e. Berry phase. As *T*/*W* decreases, clear increasing resonant mode offsets emerge in the resonant spectra measured from the Möbius- and curved-strip cavities, which is a direct evidence of the increasing Berry phase acquired in the Möbius-strip samples.

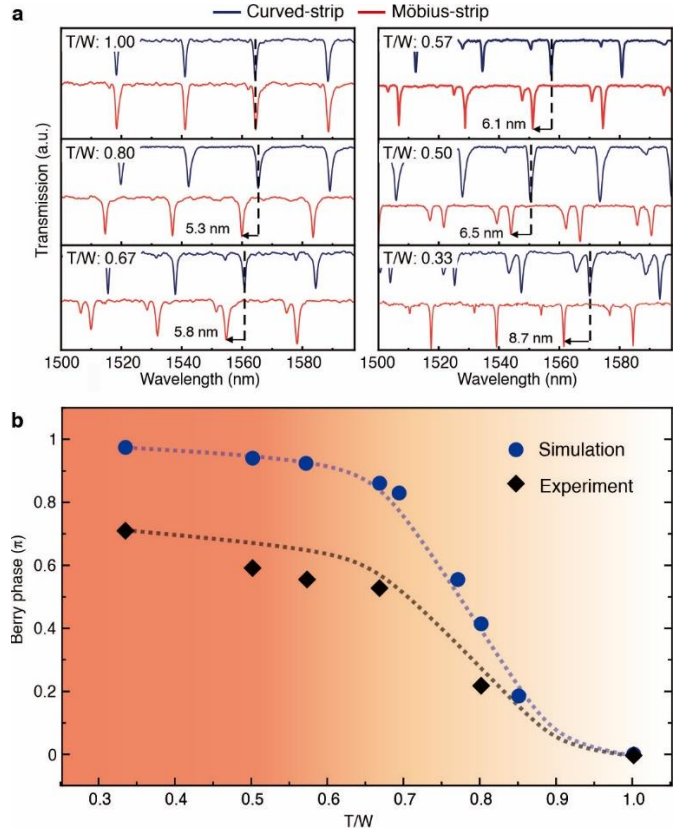

**Fig. 4 | Polarization-dependent Berry phase as a function of the cross-section geometry *T*/*W*. a,** Increasing Berry phase is revealed by comparing the resonant mode offsets between Möbius- and curved-strip cavities with decreasing *T*/*W*. **b,** Berry phase occurring in Möbius-strip cavities as a function of *T*/*W*. Solid dots and hollow squares represent data extracted from experimental and simulated spectra, respectively. Dotted lines guide the eyes.

Simulation results in Fig. 4b reveal a consistent trend and confirm that Berry phase can be freely tuned in the range from π to 0 by shaping the waveguiding cross-section of a Möbius strip. The *T*/*W* ratios ranging between ~0.65 and ~0.85 are well suited to tune Berry phase, as the ellipticity of the resonant light is particularly sensitive to *T*/*W* in that range.

The Berry phase $\varphi_B$ extracted from the resonant mode spectra is slightly lower than that derived from simulations over the whole $T/W$ ratio range as the measured resonant mode offsets are not as large as theoretical calculations. In an ideal case, the resonant mode offset is solely determined by the presence of Berry phase while Möbius- and curved-strip cavities generate the identical dynamic phase due to possession of the same size and curvature. In practical sample fabrication, a significant curvature discrepancy exists between the Möbius- and curved-strips due to the structural-symmetry-induced inherent strain, which reduces the mode offsets for the two types of cavities. As such, the deviation between experimental and simulated phase values constantly exists in the whole investigated $T/W$ ratio range except for $T/W = 1$, showing the same evolution trend of Berry phase as a function of $T/W$.

**Discussion**

The above discussion mainly focuses on circulating light with an optical electric field parallel to the strip plane (IP modes). Furthermore, resonant modes with an optical electric field perpendicular to the strip plane, termed "out-of-plane (OP)" modes, can be also well supported for the generation of Berry phase at sufficiently large values of $T$. Similar to the IP modes, OP modes with the linear and elliptical polarizations can be formed for the generation of the variable Berry phase. Simulation and experimental results of OP modes and associated Berry phases as represented in Supplementary Figs. S8 – 9 and Supplementary Notes 6 - 7 behave similarly to those of IP modes.

As a topological effect arising in cyclic adiabatic evolution, Berry phases for photons are observed and systematically investigated using all-dielectric optical Möbius-strip cavities. The variable cross-section of the waveguiding strip produces different responses of light towards the twisted topology. Curved-strip cavities with the same curvature are taken as a topologically trivial counterpart to extract Berry phase. Being able to control the cross-sectional strip dimension, the role of Berry phase in resonance modes is revealed both experimentally and theoretically, offering a guideline for manipulating the ellipticities of circulating light waves and the resultant Berry phase in the range of π to 0.

The topological robustness of Berry phase is a result of the gauge invariance in the adiabatic evolution[34]. The Möbius waveguiding structures operate in the optical wavelength range and are of micrometer size, which is much smaller than ones in the previous reports using helical waveguides and other open-light-path systems[15,16,35]. Such a miniaturized optical component is suited for a new generation of on-chip integrable systems with excellent topological robustness for both fundamental physics and practical applications. The programming of optical "Möbiosity" as a new route of light-topology shaping, has the potential to serve as a versatile knob for all-optical manipulation of both classical bits and qubits, and implies promising functionalities, such as supporting optical framed knots as information carriers[36], and quantum logic gates[37,38] in quantum computation and simulation.

**Acknowledgements**

The authors thank Haifeng Xu and Ronny Engelhard for the technical support and Suwit Kiravittaya for helpful discussions. L. B. M. acknowledges financial support by the Würzburg-Dresden Cluster of Excellence on Complexity and Topology in Quantum Matter–ct.qmat (EXC 2147, project-ID 390858490) and German Research Foundation (MA 7968/2-1). O.G.S. acknowledges financial support by the Leibniz Program of the German Research Foundation (SCHM 1298/26-1). J. W. acknowledges the support of the National Natural Science Foundation of China (Grant No. 62105080). V. M. F. acknowledges financial support by the German Research Foundation (DFG, Project No. FO 956/6-1).

**Author Contributions**

L. B. M., O. G. S., and J. W. conceived the study inspired from the initial investigation together with S. L. and V. M. F.. J. W., S. L., C. H. L., R. T., and V. M. F. performed the theoretical calculations and analysis. S. V., L. S., and M. M. S. fabricated the samples. S. V. and J. W. conducted the optical experiments. S. B. conducted SEM characterizations. J. W., L. B. M., Y. Y., and S. L. analyzed the data. J. W. and L. B. M. wrote the manuscript. All authors discussed the results and contributed to the manuscript.

**Competing financial interests**

The authors declare no competing financial interests.

## References

[1] Cohen, E., Larocque, H., Bouchard, F., Nejadsattari, F. *et al.* Geometric phase from Aharonov–Bohm to Pancharatnam–Berry and beyond. *Nat. Rev. Phys.* **1**, 437-449 (2019).
[2] Wang, J.and Zhang, S.-C. Topological states of condensed matter. *Nat. Mater.* **16**, 1062-1067 (2017).
[3] Zhang, Y., Tan, Y.-W., Stormer, H. L.and Kim, P. Experimental observation of the quantum Hall effect and Berry's phase in graphene. *Nature* **438**, 201-204 (2005).
[4] Xiao, M., Ma, G., Yang, Z., Sheng, P. *et al.* Geometric phase and band inversion in periodic acoustic systems. *Nat. Phys.* **11**, 240-244 (2015).
[5] Sonoda, H. Berry's phase in chiral gauge theories. *Nucl. Phys. B* **266**, 410-422 (1986).
[6] Datta, D. P. Geometric phase in vacuum instability: Applications in quantum cosmology. *Phys. Rev. D* **48**, 5746-5750 (1993).
[7] Yale, C. G., Heremans, F. J., Zhou, B. B., Auer, A. *et al.* Optical manipulation of the Berry phase in a solid-state spin qubit. *Nat. Photon.* **10**, 184-189 (2016).
[8] Slussarenko, S., Alberucci, A., Jisha, C. P., Piccirillo, B. *et al.* Guiding light via geometric phases. *Nat. Photon.* **10**, 571-575 (2016).
[9] Bliokh, K. Y., Rodríguez-Fortuño, F. J., Nori, F.and Zayats, A. V. Spin–orbit interactions of light. *Nat. Photon.* **9**, 796-808 (2015).
[10] Pancharatnam, S. Generalized theory of interference, and its applications. *Proc. Indian Acad. Sci.* **44**, 247-262 (1956).
[11] Berry, M. V. Quantal phase factors accompanying adiabatic changes. *Proc. R. Soc. London Ser. A* **392**, 45-57 (1984).
[12] Berry, M. V. Interpreting the anholonomy of coiled light. *Nature* **326**, 277-278 (1987).
[13] Berry, M. V. The adiabatic phase and Pancharatnam's phase for polarized light. *J. Mod. Opt.* **34**, 1401-1407 (1987).
[14] Chiao, R. Y.and Wu, Y.-S. Manifestations of Berry's Topological Phase for the Photon. *Phys. Rev. Lett.* **57**, 933-936 (1986).
[15] Tomita, A.and Chiao, R. Y. Observation of Berry's topological phase by use of an optical fiber. *Phys. Rev. Lett.* **57**, 937-940 (1986).
[16] Bliokh, K. Y., Niv, A., Kleiner, V.and Hasman, E. Geometrodynamics of spinning light. *Nat. Photon.* **2**, 748-753 (2008).
[17] Liu, Y., Guo, Q., Liu, H., Liu, C. *et al.* Circular-polarization-selective transmission induced by spin-orbit coupling in a helical tape waveguide. *Phys. Rev. Appl.* **9** (2018).
[18] Patton, R. J.and Reano, R. M. Rotating polarization using Berry's phase in asymmetric silicon strip waveguides. *Opt. Lett.* **44**, 1166-1169 (2019).
[19] Ma, L. B., Li, S. L., Fomin, V. M., Hentschel, M. *et al.* Spin-orbit coupling of light in asymmetric microcavities. *Nat. Commun.* **7**, 10983 (2016).
[20] Xu, Q., Chen, L., Wood, M. G., Sun, P. *et al.* Electrically tunable optical polarization rotation on a silicon chip using Berry's phase. *Nat. Commun.* **5**, 5337 (2014).
[21] Shao, Z., Zhu, J., Chen, Y., Zhang, Y. *et al.* Spin-orbit interaction of light induced by transverse spin angular momentum engineering. *Nat. Commun.* **9**, 926 (2018).
[22] Tanda, S., Tsuneta, T., Okajima, Y., Inagaki, K. *et al.* Crystal topology: A Mobius strip of single crystals. *Nature* **417**, 397-398 (2002).
[23] Manoharan, H. C. Topological insulators: A romance with many dimensions. *Nat. Nanotechnol.* **5**, 477-479 (2010).
[24] Greiner, J., Klose, S., Reinsch, K., Schmid, H. M. *et al.* Evolution of the polarization of the optical afterglow of the gamma-ray burst GRB030329. *Nature* **426**, 157-159 (2003).
[25] Rzepa, H. S. Möbius aromaticity and delocalization. *Chem. Rev.* **105**, 3697-3715 (2005).
[26] Fomin, V. M., Kiravittaya, S.and Schmidt, O. G. Electron localization in inhomogeneous Möbius rings. *Phys. Rev. B* **86**, 195421 (2012).
[27] Yin, Y., Li, S., Engemaier, V., Saei Ghareh Naz, E. *et al.* Topology induced anomalous plasmon modes in metallic Möbius nanorings. *Laser Photonics Rev.* **11**, 1600219 (2017).
[28] Bauer, T., Banzer, P., Karimi, E., Orlov, S. *et al.* Observation of optical polarization Möbius strips. *Science* **347**, 964-966 (2015).
[29] Song, Y., Monceaux, Y., Bittner, S., Chao, K. *et al.* Mobius Strip Microlasers: A Testbed for Non-Euclidean Photonics. *Phys. Rev. Lett.* **127**, 203901 (2021).

[30] Kreismann, J. and Hentschel, M. The optical Möbius strip cavity: Tailoring geometric phases and far fields. *Europhys. Lett.* **121**, 24001 (2018).

[31] Li, S. L., Ma, L. B., Fomin, V. M., Böttner, S. *et al.* Non-integer optical modes in a Möbius-ring resonator. *arXiv:1311.7158 [physics.optics]* (2013).

[32] Padgett, M. and Allen, L. Light with a twist in its tail. *Contem. Phys.* **41**, 275-285 (2000).

[33] Lei, F., Tkachenko, G., Ward, J. M. and Nic Chormaic, S. Complete Polarization Control for a Nanofiber Waveguide Using Directional Coupling. *Phys. Rev. Appl.* **11** (2019).

[34] Aharonov, Y. and Anandan, J. Phase change during a cyclic quantum evolution. *Phys. Rev. Lett.* **58**, 1593-1596 (1987).

[35] Chiao, R. Y., Antaramian, A., Ganga, K. M., Jiao, H. *et al.* Observation of a topological phase by means of a nonplanar Mach-Zehnder interferometer. *Phys. Rev. Lett.* **60**, 1214-1217 (1988).

[36] Larocque, H., D'Errico, A., Ferrer-Garcia, M. F., Carmi, A. *et al.* Optical framed knots as information carriers. *Nat. Commun.* **11**, 5119 (2020).

[37] Jones, J. A., Vedral, V., Ekert, A. and Castagnoli, G. Geometric quantum computation using nuclear magnetic resonance. *Nature* **403**, 869-871 (2000).

[38] Song, Y., Lim, J. and Ahn, J. Berry-phase gates for fast and robust control of atomic clock states. *Phys. Rev. Res.* **2**, 023045 (2020).

**Methods**

**Numerical simulation.** 3D numerical simulations (COMSOL Multiphysics Wave Optics module) were carried out for both Möbius and curved strips. Key parameters were designed to meet the requirement of practical fabrication procedure and operation at telecommunication wavelengths (radius $R = 10$ µm, width of 0.9 to 4 µm, and thickness of 0.6 to 1.5 µm). Models were generated using MATLAB with equations provided in Supplementary Note 1. Optical mode fields were simulated in the wavelength range from 1500 to 1560 nm. IP and OP excitations were conducted by a local line current source at the waveguide core oscillating along the corresponding directions. The local polarization state was extracted by analyzing the electric field distributions at the field maxima point of the waveguide's cross-section spanning from one mode antinode to node and presented using a projected polar plot.

**Device fabrication.** Pure quartz glass substrates were cleaned by immersion in acetone and isopropyl alcohol, ultrasonicated (Elmasonic S, Elma Schmidbauer GmbH) and dried with $N_2$ gas. The microcavities were fabricated using 3D direct laser writing (DLW, Photonic Professional GT, Nanoscribe GmbH) on the cleaned quartz substrates. The laser lithography photoresist was drop-casted onto the substrates. The principle of DLW system is based on multi-photon lithography and 3D scanning of focused laser beam within the sample volume. A high-resolution galvanometer mirror system is used for the laser beam scanning in the x- and y- plane, which allows for positioning each volumetric pixel with an accuracy of 10 nm. The scanning is controlled by the DeScribe software in Nanoscribe which is also used to design the structures. During the fabrication process, the negative photoresist (IP-Dip, Nanoscribe GmbH) is polymerized by two-photon absorption (@780 nm). After the exposure, the structures were developed in mr-Developer (mr-Dev 600, Micro Resist Technology GmbH) followed by rinsing in isopropanol which resulted in fabricated structures bonded on the substrate. After that, the structures were dried using critical point drier (Autosamdri-931, Tousimis Research Corporation) to remove the solvent on the structures. Each of the fabricated Möbius- and curved-strip microstructures is supported by two ca. 20 µm high pylons to avoid optical leakage to the substrate.

**Device characterization.** For SEM imaging the samples were sputter coated with a thin (ca. 8 nm) Cr layer to avoid charging and to improve the contrast. The SEM images were taken in an NVision40 (Carl Zeiss GmbH Oberkochen) using a primary beam energy of 5 keV. Optical transmission measurements were conducted by an evanescently coupled tapered fiber. The optical setup and details of preparing tapered fiber are included in Supplementary Note 2. A wavelength-tunable infrared laser (Tunics 100S-HP, Yenista) was used as a light source with a scanning range covering 1500-1600 nm. The transmission signal was measured using an InGaAs switchable-gain photodetector (PDA20CS-EC, Thorlabs). Resonances with a maximized extinction ratio of > 10 dB were obtained by fine adjusting the coupling gap spacing. The polarization state of the incident laser light was controlled by a fiber-based three-paddle polarization controller. Launching of linear-polarized light with tilted orientation was carefully examined by connecting a fiber-coupled polarimeter (PAX1000IR2, Thorlabs) at the output. Polarization mapping was carried out by adjusting the polarization angle of the input laser light in a step of 10°. Berry phase was estimated by deriving the compensated phase value (i.e., resonant wavelength offset) for spectral matching of resonance modes in Möbius- and curved-strip cavities with identical structural parameters.

**Data availability**

The main data supporting the findings of this study are available within the Supplementary Information. Additional data are available from the corresponding authors upon reasonable request.